\documentclass[aip, rsi, reprint, amssymb, amsfonts, groupedaddress]{revtex4-1}

\pdfoutput=1

\usepackage{amsmath}
\usepackage{bm}
\usepackage{graphicx}
\usepackage[unicode]{hyperref}
\usepackage[latin1]{inputenc}
\usepackage{array}
\usepackage{upgreek}
\usepackage{color}
\usepackage{booktabs}

\begin{document}

\title{Vibration isolation with high thermal conductance for a cryogen-free dilution refrigerator }

\author{Martin de Wit} \thanks{Both authors contributed equally to this work}
\author{Gesa Welker} \thanks{Both authors contributed equally to this work}
\author{Kier Heeck}
\author{Frank M. Buters}
\author{Hedwig J. Eerkens}
\author{Gert Koning}
\author{Harmen van der Meer}
\author{Dirk Bouwmeester}
\altaffiliation[Also at ]{Department of Physics, University of California, Santa Barbara, California 93106, USA.}
\author{Tjerk H. Oosterkamp}
\email{oosterkamp@physics.leidenuniv.nl}
\affiliation{Leiden Institute of Physics, Leiden University, PO Box 9504, 2300 RA Leiden, The Netherlands}

\date{\today}

\begin{abstract}
We present the design and implementation of a mechanical low-pass filter vibration isolation used to reduce the vibrational noise in a cryogen-free dilution refrigerator operated at 10 mK, intended for scanning probe techniques. We discuss the design guidelines necessary to meet the competing requirements of having a low mechanical stiffness in combination with a high thermal conductance. We demonstrate the effectiveness of our approach by measuring the vibrational noise levels of an ultrasoft mechanical resonator positioned above a SQUID. Starting from a cryostat base temperature of 8 mK, the vibration isolation can be cooled to 10.5 mK, with a cooling power of 113 $\upmu$W at 100 mK. We use the low vibrations and low temperature to demonstrate an effective cantilever temperature of less than 20 mK. This results in a force sensitivity of less than 500 zN/$\mathrm{\sqrt{Hz}}$, and an integrated frequency noise as low as 0.4 mHz in a 1 Hz measurement bandwidth.
\end{abstract}

\maketitle

\section{Introduction}

The ability to work at very low temperatures with minimal mechanical environmental noise is vital for many scanning probe techniques which depend heavily on a stable tip-sample seperation, such as Scanning Tunneling Microscopy (STM) \cite{pan1999,allworden2018} and Magnetic Resonance Force Microscopy (MRFM) \cite{degen2009,wagenaar2016}, as well as when investigating the quantum properties of macroscopic objects where resonators with extremely low mode temperatures are required \cite{kleckner2006,bassi2013}. Most of these experiments are done in "wet" cryostats that use a liquid helium bath to reach low temperatures without adding too much mechanical vibrations \cite{kuhn2014}. However, with the decreasing availability of helium, many new setups are now based on "dry" cryostats, where the liquid helium bath is replaced by a cryocooler. The downside of these so called cryogen-free cryostats is that the cryocooler introduces significant vibrations \cite{chijioke2010,olivieri2017}, which have to be decoupled from the experiment.

The challenge for these experiments is that the technical requirements to cool an experiment to temperatures well below 1K are often conflicting with those necessary to achieve very low mechanical vibrations. To obtain the lowest possible temperature one should have good thermal conduction by having a rigid contact between the experiment and the source of cooling. However, in order to reduce mechanical vibrations to acceptable levels, the connection between the two should be a soft as possible, for instance by suspending the experiment from springs. To tackle this design conundrum, often a global solution is used, where the entire instrument is isolated, typically by making it very heavy. The actual experiment is then connected as stiff as possible to the cold stage in order to get a good heat conductance and to minimize the displacement noise due to residual vibrations \cite{misra2013,song2010,singh2013,allworden2018}. This approach works particularly well when the main sources of noise originate from outside of the cryostat, which is often the case for wet cryostats, but does not necessarily apply to dry systems. For dry systems, where most of the noise is generated at the cryostat, a local solution is required, where both the issues of vibration isolation and thermal conductance have to be solved by a combined solution within the cryostat\cite{foley2004,caparrelli2006,pelliccione2013,haan2014}. This often leads to a compromise for one of the two. Here we present a design which optimizes both aspects.

The vibration isolation presented here is intended to be used for a low temperature MFM/MRFM setup, where an ultrasoft resonator is used to measure the properties of various spin systems\cite{poggio2010}. Due to the low stiffness and high quality factor of the resonator the system is extremely sensitive to small forces\cite{vinante2012}, but also to vibrations. Therefore, we will demonstrate the effectiveness of the vibration isolation by analyzing the displacement noise spectrum and thermal properties of the resonator, showing that our vibration isolation allows us to perform highly sensitive measurements in a cryogen-free dilution refrigerator. We will start by explaining the correspondence between electrical and mechanical networks, as this analogue proves to be very useful for calculating the optimal design of the mechanical filter.

\section{Filter design}

Commonly, the development of mechanical vibration isolation relies heavily on finite element simulations to determine the design parameters corresponding to the desired filter properties. In these simulations, the initial design is tweaked until the desired filter properties are found. Instead, we determined the parameters of our mechanical filter by first designing an electrical filter with the desired properties, and then converting this to the mechanical equivalent using the current-force analogy between electrical and mechanical networks. This allows us to precisely specify the desired filter properties beforehand, from which we can then calculate the required mechanical components. We therefore find the optimal solution using analytical techniques rather than using complex simulations. As we will see later, this also allows us to adopt our design principle across many frequency scales without requiring a new finite element analysis.
The corresponding quantities for the analogy between electrical and mechanical circuits can be found in Table. \ref{table:table}. We choose the current-force analogy over the voltage-force analogy \cite{fowles2005} because the former conserves the topology of the network.

\begin{table}[]
	\begin{tabular}{@{}p{2.2cm} p{1.8cm}|p{2.2cm} p{1.8cm}@{}}
		\multicolumn{2}{c|}{\textbf{Electrical}} 							& \multicolumn{2}{c}{\textbf{Mechanical}} \\ \midrule \hline
		\textbf{Variable}   & \textbf{Symbol}   							& \textbf{Variable}    & \textbf{Symbol}   \\
		Current             & I {[}A{]}         							& Force                & F {[}N{]}         \\
		Voltage             & U {[}V{]}         							& Velocity             & v {[}m/s{]}       \\
		Impedance           & Z {[}$\Omega${]}     							& Admittance           & Y {[}s/kg{]}      \\
		Admittance          & Y {[}1/$\Omega${]}    						& Impedance            & Z {[}kg/s{]}      \\
		Resistance			& R {[}$\Omega${]}								& Responsiveness	   & 1/D {[}s/kg{]}	   \\
		Inductance          & L {[}H{]}         							& Elasticity           & 1/k {[}m/N{]}     \\
		Capacitance         & C {[}F{]}         							& Mass                 & m {[}kg{]}        \\ \midrule \hline
		\multicolumn{2}{c|}{\includegraphics{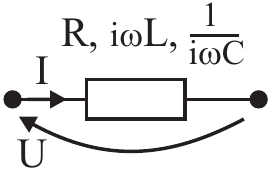}}            & \multicolumn{2}{c}{\includegraphics{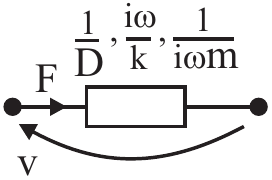}}                   
	\end{tabular}
	\caption{Table of corresponding electrical and mechanical quantities.}
	\label{table:table}
\end{table}

\begin{figure}
	\centering
	\includegraphics[width=\columnwidth]{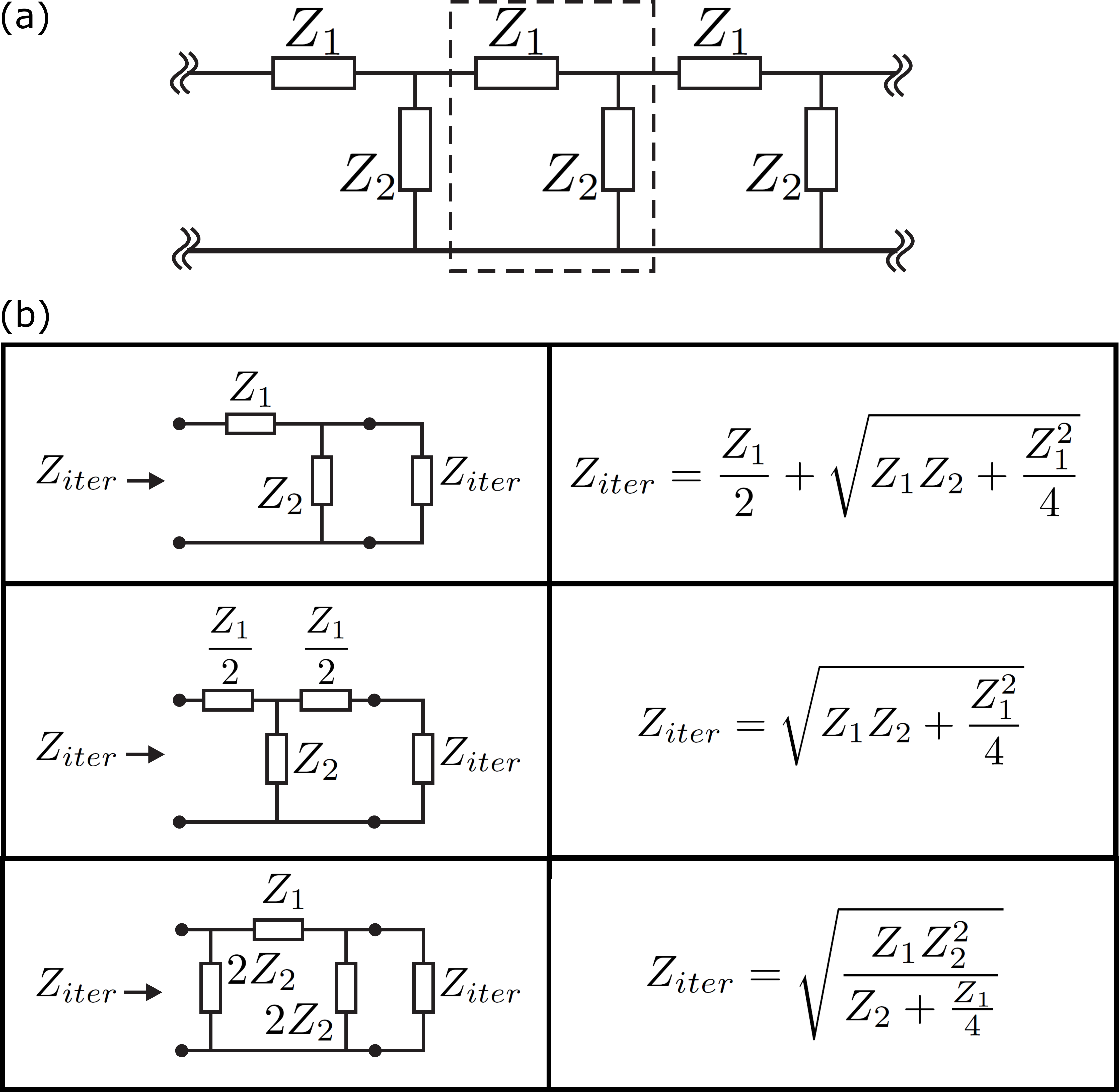}
	\caption{a) General scheme of a filter composed of identical sections, with one unit cell indicated by the black dotted box. b) Three options for the design of the unit cell for an LC filter, with on the right the corresponding values for the iterative impedance $Z_{iter}$.}
	\label{figure:Campbell}
\end{figure}

To design our desired filter, we follow the method of Campbell for the design of LC wave-filters \cite{campbell1917}. Campbell's filter design method is based on two requirements:
\begin{itemize}
\item The filter is thought to be composed out of an infinite repetition of identical sections, as shown in Fig. \ref{figure:Campbell}(a), where a single section (also called unit cell) is indicated by the black dotted box.
\item The sections have to be dissipationless to prevent signal attenuation in the passband. Therefore the impedances of all elements within the section have to be imaginary.\\
\end{itemize} 
Following these requirements, the edges of the transmitted frequency band of the filter are defined by the inequality
\begin{equation}\label{eq:BW}
	-1 \leq \frac{Z_1}{4 Z_2} \leq 0
\end{equation}
The iterative impedance is the input impedance of a unit cell when loaded with this impedance. In order to prevent reflections within the pass-band, the signal source and the load should have internal impedances equal to $Z_{iter}$. The iterative impedance should be real and frequency-independent, because this maximizes the power transfer within the pass-band and is easiest to realize.

There are three principle choices for the unit cell, all given in Fig. \ref{figure:Campbell}(b). The total attenuation is determined by the number of unit cells. Each unit cell acts like a second order filter, adding an extra 40 dB per decade to the high frequency asymptote of the transfer function. This attenuation is caused by reflection, not by dissipation, which is very important for low-temperature applications.

\begin{figure}
	\centering
	\includegraphics[width=\columnwidth]{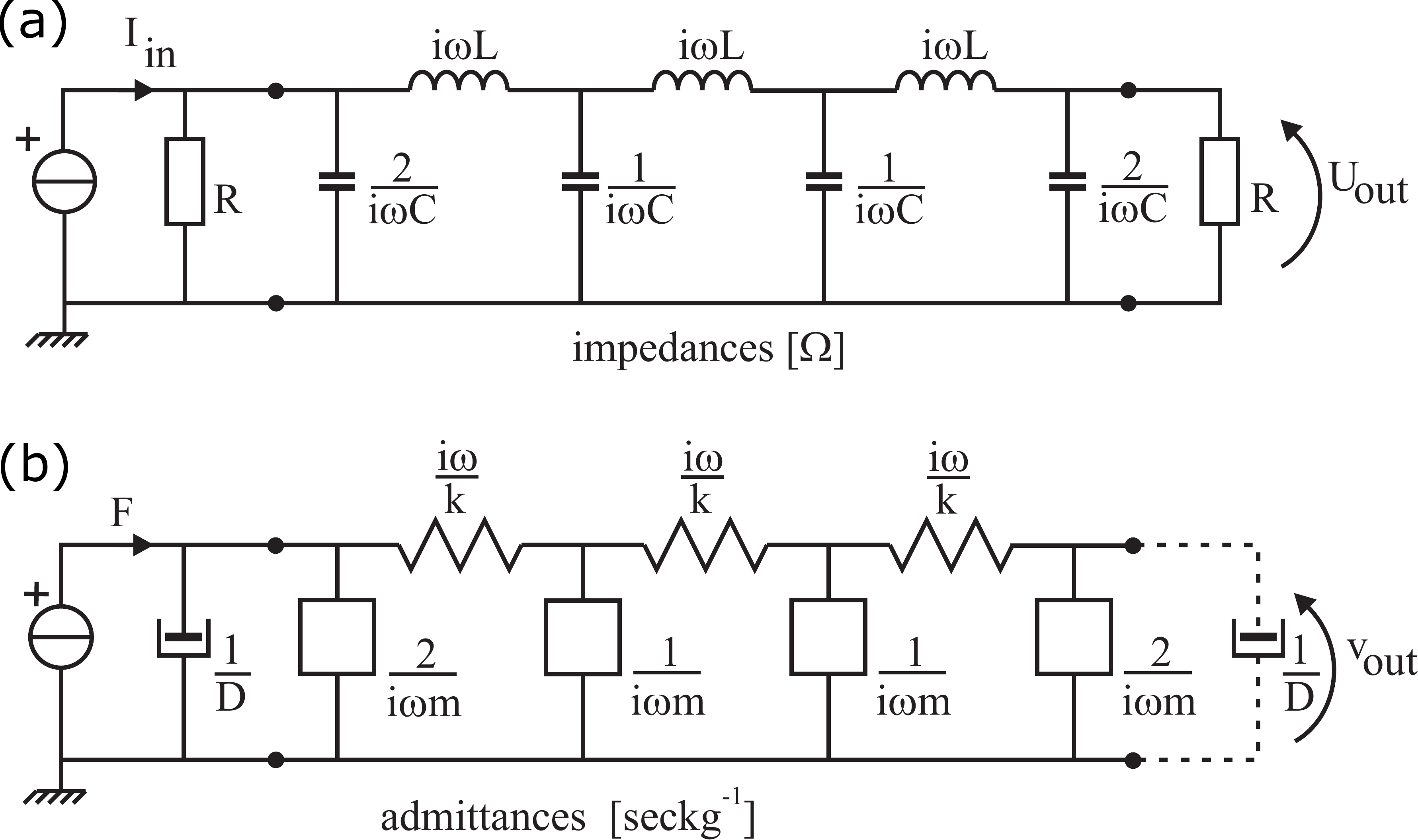}
	\caption{a) Electrical filter consisting of two unit cells. b) Corresponding mechanical filter.}
	\label{figure:elmech}
\end{figure}

The design of the mechanical filter is straightforward when we use the third option from Fig. \ref{figure:Campbell}(b) with $Z_1 = \frac{1}{Y_1} = i \omega L$, and $Z_2 = \frac{1}{Y_2} = \frac{1}{i \omega C}$. The resulting electrical low-pass filter is shown in figure \ref{figure:elmech}(a). Please note that the two neighboring $2 Z_2$ in the middle add up to $Z_2$. We can use Eq. \ref{eq:BW} to calculate the band edges: $\omega_1 = 0$ and $\omega_2 = \frac{2}{\sqrt{LC}}$.

With the electrical filter figured out, we make the transfer to the mechanical filter according to the correspondence as outlined in Table. \ref{table:table}. As the electrical inductance corresponds to mechanical elasticity, the coils are replaced by mechanical springs with stiffness $k$. The capacitors are replaced by masses in the mechanical filter. Note that the first mass has the value $\frac{m}{2}$ due to the specific unit cell design. The current source becomes a force source and the electrical input and load admittances become mechanical loads (dampers). The final mechanical circuit is depicted in Fig. \ref{figure:elmech}(b). Going to the mechanical picture also implies a conversion between impedance and admittance in Eq. \ref{eq:BW}:
\begin{equation}\label{eq:BWmech}
	-1 \leq \frac{Y_1}{4 Y_2} \leq 0
\end{equation}
With $Y_1 = \frac{k}{i\omega}$ and $Y_2 = i \omega m$, this leads to the band edges $\omega_1 = 0$ and $\omega_2 = 2 \sqrt{\frac{k}{m}}$.

We now have a design for the unit cell of a general mechanical low-pass filter. The bandwidth and corner frequency are determined by the choice of the stiffness $k$ and mass $m$, which can be taylored to the needs of a specific experimental setup. In practice, only corner frequencies between a few Hz and 50 kHz can be easily realized. At too low frequencies, the necessary soft springs will not be able to support the weight anymore, whereas above 50 kHz, the wavelength of sound in metals comes into play, potentially leading to the excitation of the eigenmodes of the masses.

\begin{figure}
	\centering
	\includegraphics[width=\columnwidth]{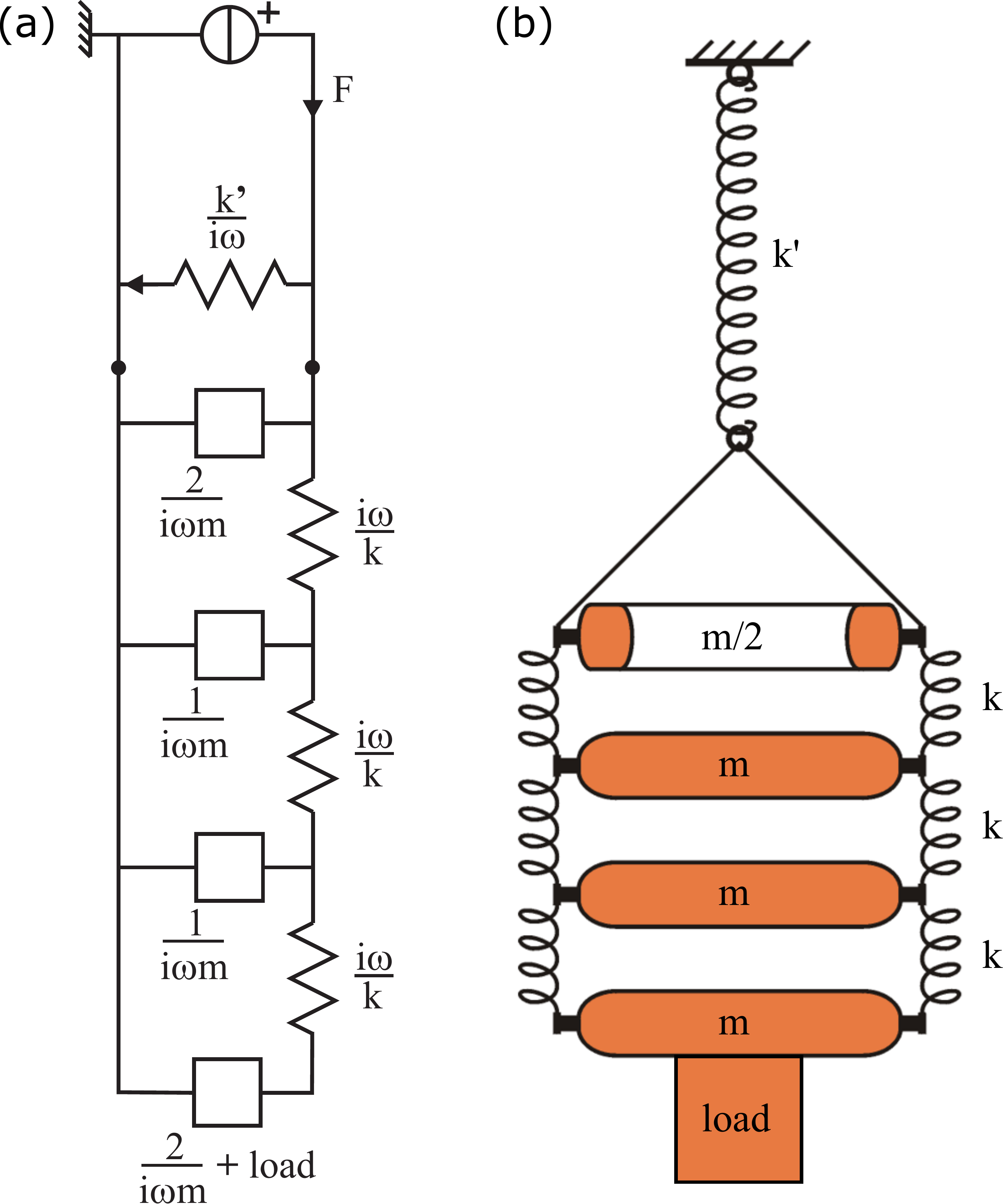}
	\caption{(a) Circuit diagram and (b) schematic overview of the mechanical low-pass filter based on the outlined theory. Note that the damper at the input is missing.}
	\label{figure:LPF_Theory}
\end{figure}

A second practical challenge is the realization of the damper at the end of the filter. It should be connected to the mechanical ground, just as an electrical load is connected to the electrical ground. This is, however, not possible, because this mass reference point is defined by earth's gravity. The alternative to a damper as real-valued load is using a purely reactive load: more mass. Simulations show that adding mass to the $m/2$ of the filters last mass does not significantly alter the frequency characteristics of the filter, and even increases the attenuation. There is no strict limit on the weight of the added mass. In fact, adding more will in principle improve the filter. In practice, the limit depends on the choice of springs, which should be able to carry the weight whilst staying in the linear regime. The downside of replacing the damper with mass is that we lose the suppression of the resonance frequencies of the filter. We have chosen a final mass with a weight equal to the previous mass. The circuit diagram and schematic for the final design of the mechanical low-pass filter is shown in Fig. \ref{figure:LPF_Theory}. Note that the damper at the input is missing, for experimental reasons which will be explained in Sec. \ref{sec:VibExp}.

\section{Practical design and implementation} 
Our setup is based on a Leiden Cryogenics CF-1400 dilution refrigerator with a base temperature of 8 mK and a measured cooling power of 1100 $\upmu$W at 120 mK. Various vibration damping systems have been implemented previously, including mechanically decoupling the two-stage pulsetube cryocooler from the cryostat, and suspending the bottom half of the cryostat from springs between the 4K-plate and the 1K-plate. These adjustments are described in detail by \citeauthor{haan2014}\cite{haan2014}. In the current paper, we focus only on the implementation and performance of the mechanical low pass filter below the mixing chamber.

\begin{figure*}
	\centering
	\includegraphics[width=\textwidth]{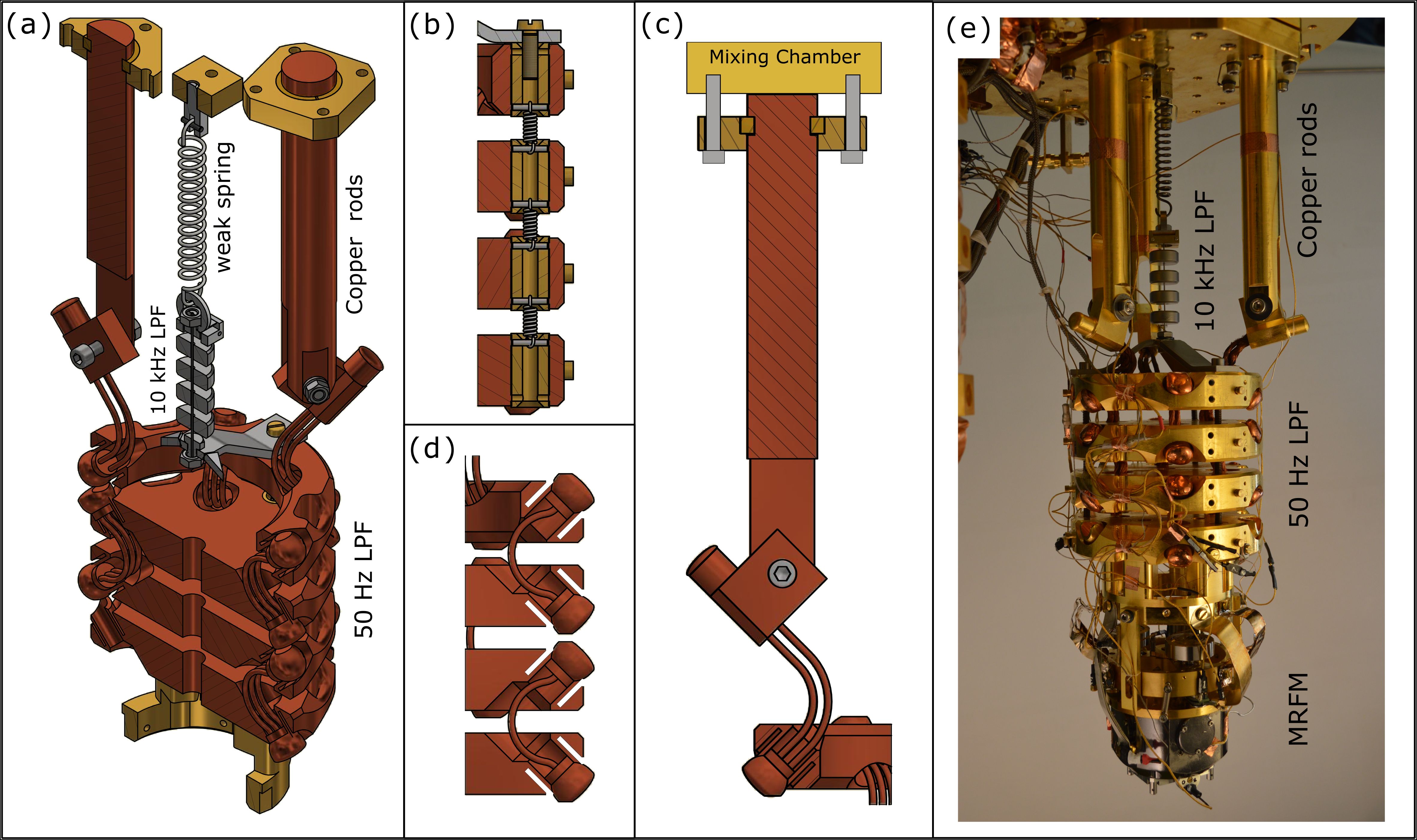}
	\caption{(a) Schematic drawing of the full design of the low temperature vibration isolation. It consists of a weak spring, a 10 kHz low-pass filter and a 50 Hz low-pass filter. The full length of the assembly is about 50 cm. The experiment can be mounted below the bottom mass. (b) Detailed schematic of the springs interconnecting the masses. The design is such that the springs can be replaced even after the filter is fully assembled and welded. (c) Detailed schematic of the thermal connection between the mixing chamber and the top mass. To get as little interfacial thermal resistance as possible, the copper rods are pressed directly against the mixing chamber. (d) Detailed schematic of the heatlinks interconnecting the masses. Of particular importance are the notches that concentrate the heat during the welding of the heatlinks. The heatlinks consists of three soft braided strands of copper. (e) Photo of the vibration isolation mounted on the mixing chamber of the dilution refrigerator.}
	\label{figure:setup}
\end{figure*}

The design of the vibration isolation based on the theory outlined in the previous section can be seen in Fig. \ref{figure:setup}(a). The isolation consists of three distinct parts: the weak spring intended to carry the weight, the 50 Hz low-pass filter acting as the main vibration isolation filter, and an additional 10 kHz low-pass filter used to remove mechanical noise from the cold head of our pulsetube at 24 kHz. 

\begin{figure}
	\centering
	\includegraphics[width=\columnwidth]{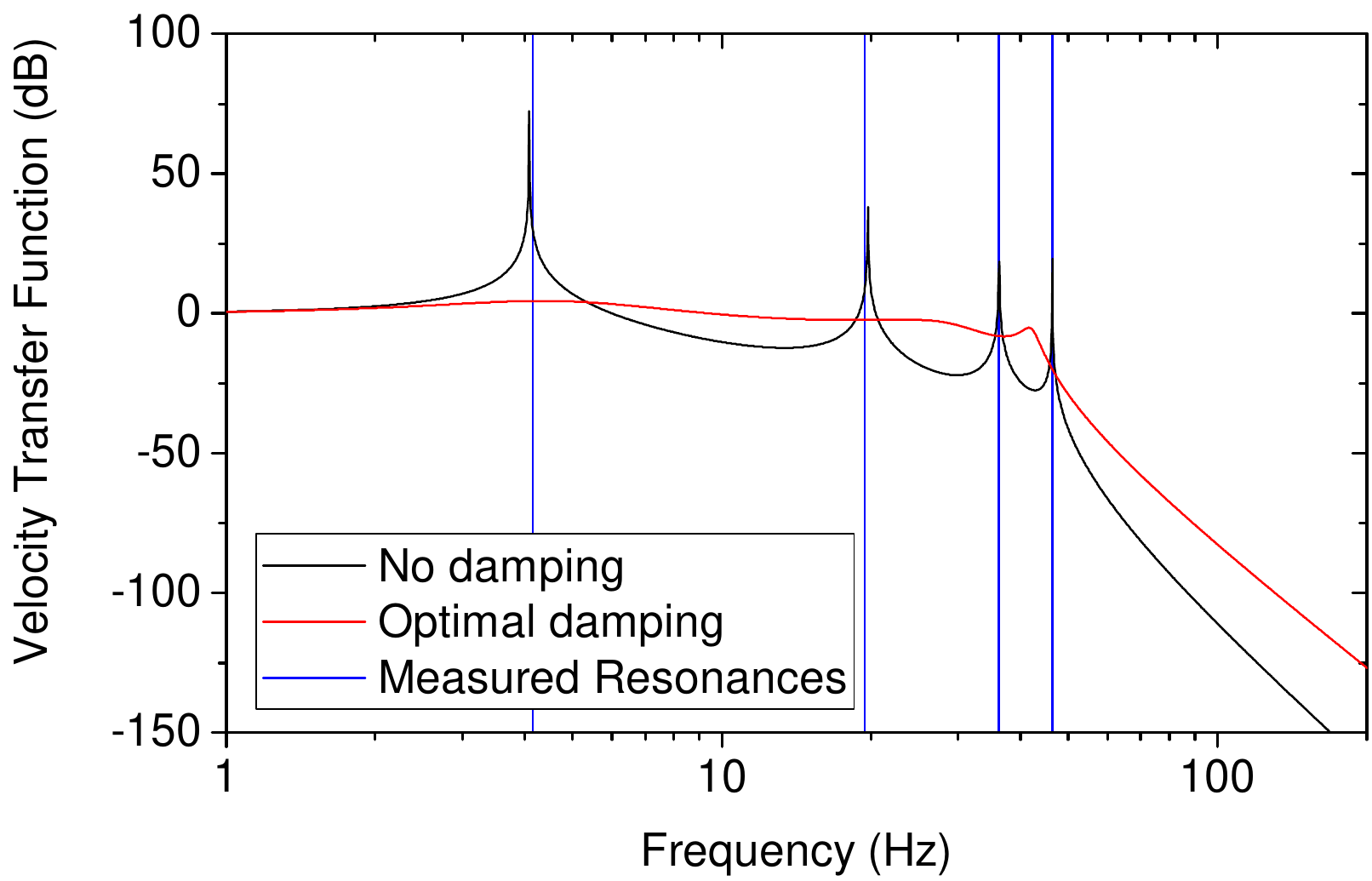}
	\caption{Theoretical velocity transfer function of the mechanical low-pass filter, calculated without damping (black line) and with optimal damping (red line). The vertical blue lines indicate the positions of the measured room-temperature resonance frequencies.}
	\label{figure:TF}
\end{figure}

The 50 Hz filter consists of 4 separate gold-plated copper masses, each connected by 3 springs. The top mass has half the weight of the other three masses, as dictated by the theory. A variation in the masses of the different plates of up to 20 \% is allowed without a significant reduction in the isolating performance. As it might be desirable to tune the internal frequencies of this mass-spring system away from mechanical vibration frequencies of the cryostat, the springs, made of stainless steel, are fully modular and can easily be replaced even after assembly, as can be seen in Fig. \ref{figure:setup}(b). When multiple springs are used at each stage, the stiffness of all springs should be as equal as possible.  In our design we have chosen a mass $m$ = 2 kg, and springs with a stiffness $k$ = 16 kN/m, leading to a combined stiffness of 48 kN/m. This choice leads to a corner frequency of $\omega_c / (2 \pi) = 50$ Hz. We have chosen to use 3 filter stages as this should give sufficient attenuation above 100 Hz. The internal resonance frequencies of this filter have been measured at room temperature by applying a driving force at the top mass of the filter and using geophones to measure the response at the bottom mass. The frequencies match well with the resonance frequencies obtained from the theoretical model, as can be seen in Fig. \ref{figure:TF}. The good agreement between theory and experiment in terms of the resonance frequencies gives confidence to also trust the model regarding the reduction of vibrations, where we expect over 100 dB of attenuation above 100 Hz. This level of attenuation is sufficient for our application with a resonator at a frequency of 3 kHz, but it is also possible to attain a larger attenuation at lower frequencies, as indicated in the previous section. The internal resonances can be suppressed by adding a well-designed damper, as demonstrated by the calculation shown as the red line in Fig. \ref{figure:TF}.

The 10 kHz filter consists of a stainless steel wire with a diameter of 1.0 mm connecting 4 stainless steel masses weighing 20 g each. The design of this second filter is also based on the previously outlined theory, just like the 50 Hz filter. This filter is necessary to remove noise that can drive high frequency internal filter modes, e.g. resonances of the masses. Once again, the theoretical internal resonances were verified experimentally, indicating that the electrical-to-mechanical filter correspondence holds for a wide range of frequencies.

Concerning the weak spring: we have chosen a stainless steel spring with length 100 mm and a spring constant of about 10 kN/m, leading to a resonance frequency of 4 Hz. However, it must be noted that this choice is not at all critical. A wide range of spring constants is allowed, as long as the weak spring can really be considered weak with regard to the springs interconnecting the masses. If a damper is added to the system, it should be in parallel to the weak spring. Please note that no additional damping is necessary in parallel to the springs between the masses in order to damp all four resonances.

When mounting the experiment including its electrical wiring, care needs to be taken to attach each wire firmly to each of the masses. Otherwise, the wires create a mechanical shortcut, thereby reducing the efficiency of the vibration isolation.

In order to be able to cool the experiments suspended from the vibration isolation to temperatures as close to the temperature of the mixing chamber as possible, we have taken great care to maximize the thermal conductance. Since the biggest bottlenecks in the thermal conductance are the stainless steel weak spring and 10 kHz low-pass filter, we bypass these components by using three solid copper rods in parallel to the weak spring, each with a diameter of 25 mm and 175 mm length, which are connected to the top mass via three soft braided copper heatlinks. We are allowed to make this thermal bypass as long as the combined stiffness of the soft heatlinks and the weak spring remains low compared to the stiffness of the interconnecting springs. The soft braided copper heatlinks consist of hundreds of intertwined copper wires with a diameter of 0.1 mm. Using a bundle of thin wires leads to a much lower mechanical stiffness than when using a single thick wire. In order to avoid a large contact resistance between the mixing chamber and the copper rods, the rods are gold-plated and placed directly against the mixing chamber plate of the dilution refrigerator. All contact surfaces are cleaned by subsequently using acetone, ethanol, and isopropanol to remove organic residue, which can reduce the thermal conductance. A strong mechanical contact is achieved using the system shown in Fig. \ref{figure:setup}(c). All clamping contacts using bolts contain molybdenum washers, as these will increase the contact force during cooldown due to the low thermal contraction coefficient of molybdenum compared to other metals.

All masses are interconnected via three sets of three soft braided copper heatlinks which are TIG welded into the masses in an argon atmosphere to prevent oxidation. The welding of the copper was made possible by the notched structure of the welding joints in the masses (see Fig. \ref{figure:setup}(d)), which are intended to concentrate the heat during welding. The gold plating was removed from the welding joint prior to the welding to prevent diffusion of the gold into the heatlinks, which would reduce the thermal conductance. The experiment is rigidly attached to the bottom mass, which should now function as a cold and vibration-free platform.

\section{Experimental results}
To characterize the performance of the vibration isolation, we install a very soft cantilever (typically used for MRFM experiments) below the bottom mass. The cantilever has a spring constant $k_0$ = 70 $\upmu$N/m, a resonance frequency $f_0$ of about 3009 Hz, and a quality factor $Q_0$ larger than 20000 at low temperatures. A magnetic particle (radius $R_0$ = 1.7 $\upmu$m) is attached to the end of the cantilever. We then compare two situations: In one configuration, the vibration isolation is operating as intended and as decribed in the previous section. In the other configuration, the vibration isolation was disabled by using a solid brass rod to create a stiff connection between the mixing chamber and the last mass of the vibration isolation. This simulates a situation where the experiment is mounted without vibration isolation.The vibrations of the setup are determined by measuring the motion of the cantilever using a SQUID \cite{usenko2011}, which measures the changing flux due to the motion of the particle. The sensitivity of this vibration measurement is limited by the flux noise of the SQUID, which can be converted to a displacement noise using the thermal motion of the cantilever and the equipartition theorem \cite{vinante2012}. We start by demonstrating the thermal properties of the vibration isolation.

\subsection{Thermal conductance}

\begin{figure*}
	\centering
	\includegraphics[width=\textwidth]{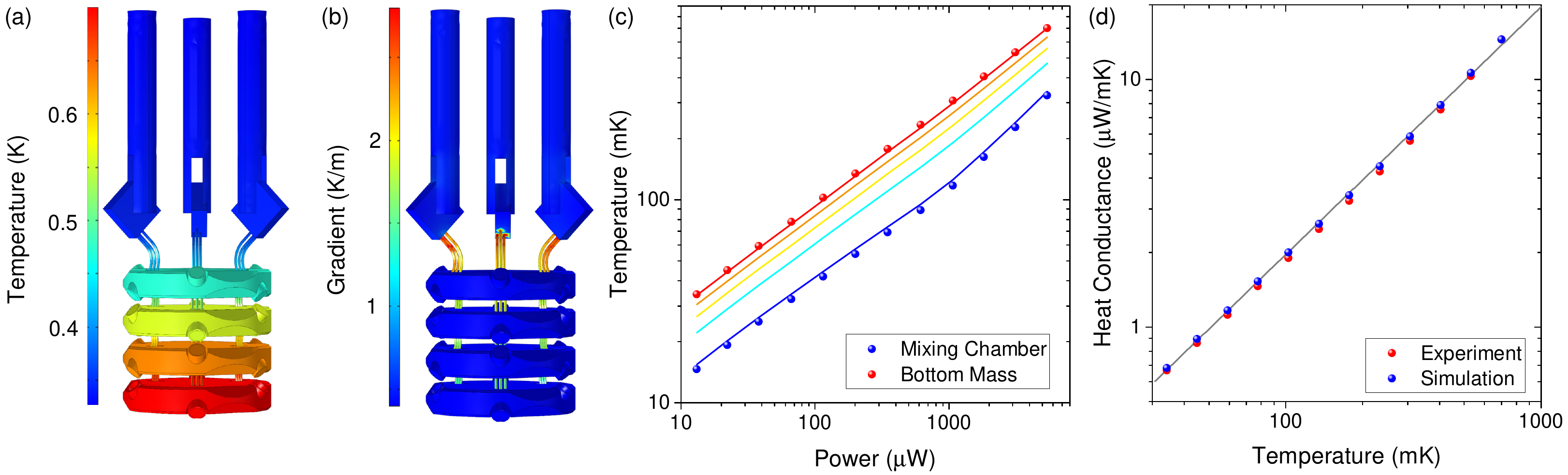}
	\caption{Measurements and finite element simulations of the thermal properties of the vibration isolation. A power is applied to the bottom mass, and the temperature of the bottom mass and the mixing chamber are measured. In the simulation, we insert the power and mixing chamber temperature, and calculate the corresponding temperature of the bottom mass to check the model. Results of the simulation for the (a) temperature and (b) temperature gradient are shown for a power of 5.4 mW. (c) Measured temperature of the mixing chamber and bottom mass as a function of the applied power. The solid lines are the simulated temperatures at each of the masses (red is the bottom mass, blue the bottom of the copper rod). At 100 mK, we find a cooling power of 113 $\upmu$W at the bottom mass. (d) Heat conductance between the bottom mass and the mixing chamber as a function of the temperature of the bottom mass. The solid line is a linear fit to the data.}
	\label{figure:K_Mass3}
\end{figure*}

To verify the effectiveness of the thermalization, we have measured the heat conductance of our vibration isolation. For the base temperature of our cryostat, which is a mixing chamber temperature of approximately 8 mK, we find that the bottom mass of the vibration isolation saturates at 10.5 mK. This already indicates a good performance of the thermalization. We then use a heater to apply a known power to the bottom mass, while we again measure the temperature of the bottom mass and the mixing chamber. This allows us to quantify an effective cooling power at the bottom mass (defined as the maximum power that can be dissipated to remain at a set temperature). At 100 mK, we measure a cooling power of 113 $\upmu$W, which is significantly higher than that of comparable soft low temperature vibration isolations described in literature \cite{moussy2001,pelliccione2013}, and only about a factor of 7 lower than the cooling power of the mixing chamber of the dilution refridgerator at the same temperature.

The experimental data is compared to a finite element simulation using Comsol Multiphysics to determine the limiting factors in the heat conductance. The results of this analysis and the experimental data are shown in Fig. \ref{figure:K_Mass3}. We use a thermal conductivity that is linearly dependent on temperature as expected for metals \cite{pobell1996}, given by $\kappa = 145 \cdot T$. The proportionality constant of \mbox{145 Wm$^{-1}$K$^{-2}$} corresponds to low purity copper \cite{woodcraft2005}. The simulated temperature distribution (for an input power of 5.4 mW) is shown in Fig. \ref{figure:K_Mass3}(a). The uniformity of the color of the masses indicates that the heatlinks interconnecting the masses are the limiting thermal resistance, something that becomes even more apparent from the plotted thermal gradient as shown in Fig. \ref{figure:K_Mass3}(b).

There is a good correspondence between the simulation and the experimental values for all applied powers, as shown in Fig. \ref{figure:K_Mass3}(c). Similar agreement is found when plotting the heat conductance between the bottom mass and the mixing chamber as a function of the temperature of the bottom mass (Fig. \ref{figure:K_Mass3}(d)).  The assumption that the heat conductivity is linearly dependent on the temperature seems to hold over the full temperature range. As the model does not include contact resistance or radiation, but only the geometry and thermal properties of the copper, we can conclude that the thermal performance of the vibration isolation is limited purely by the thermal conductance of the braided copper. Furthermore, we do not expect that other sources of thermal resistance follow this particular temperature dependence \cite{pobell1996}. So, the argon-welded connections appear to be of sufficient quality not the hinder the conductance. The performance can be improved further by making the heatlinks out of copper with a higher RRR value, and thereby a higher thermal conductivity.

\subsection{SQUID vibration spectrum} \label{sec:VibExp}
\begin{figure}
	\centering
	\includegraphics[width=\columnwidth]{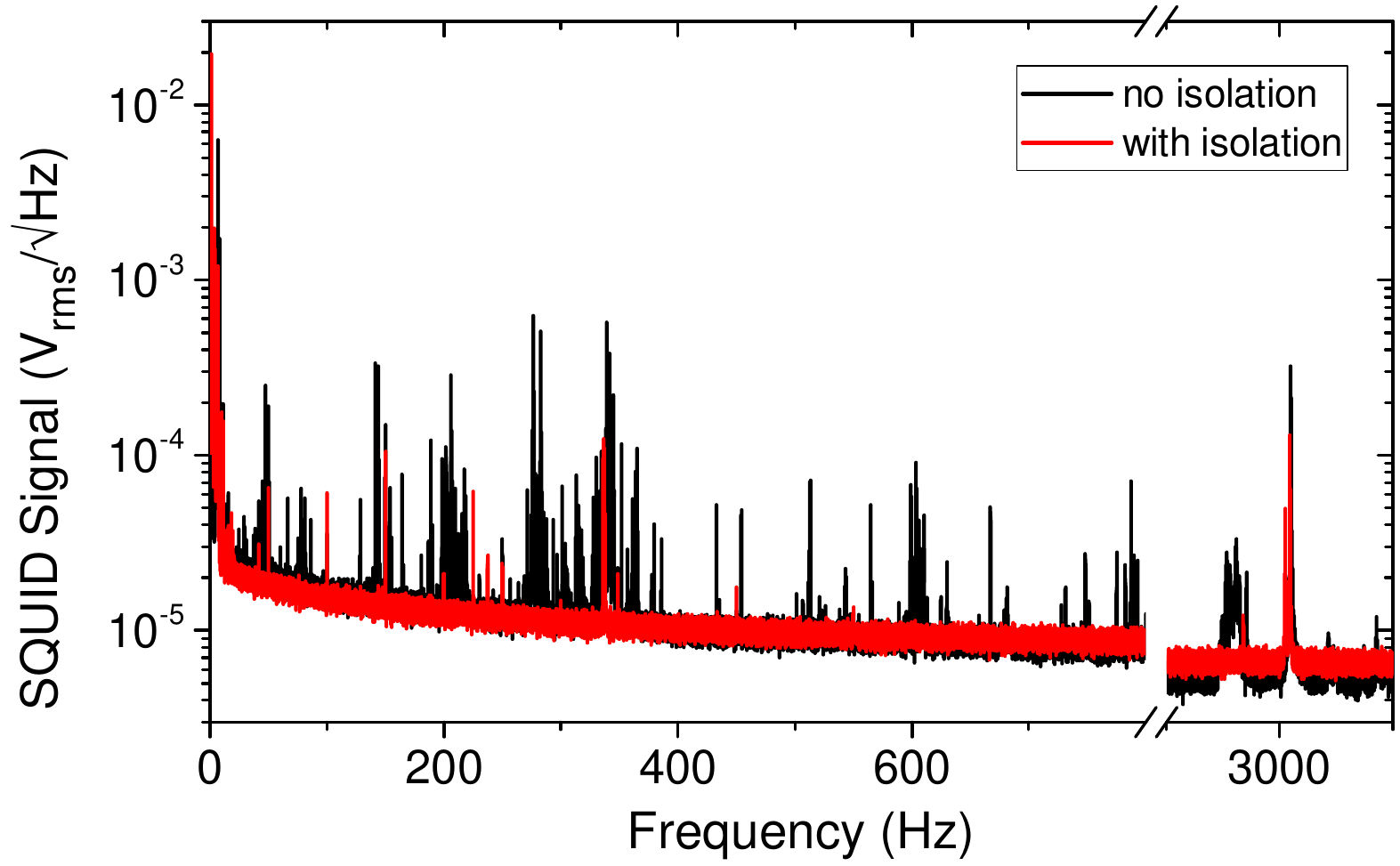}
	\caption{SQUID spectra $\sqrt{S_V}$ of the vibration noise measured at temperatures below 25 mK. The black data shows the SQUID signal with the vibration isolation disabled using the brass rod, while in red we see the measured spectrum with proper vibration isolation.}
	\label{figure:spectrum}
\end{figure}

The performance of the vibration isolation is shown in Fig. \ref{figure:spectrum}, where we plot the measured SQUID spectra for the two different situations: In the red data, the vibration isolation is in full operation. The black data shows the situation when the vibration isolation is disabled. A clear improvement is visible for nearly all frequencies above 5 Hz. We focus on the region between 0 and 800 Hz to indicate how effective almost all vibrations are reduced to below the SQUID noise floor, and on the region around 3009 Hz as this is the resonance frequency of our cantilever. The conversion factor ($c$)between SQUID voltage and displacement is about 0.78 mV/nm for the black spectrum, and 0.56 mV/nm for the red spectrum, where the small difference is caused by a slightly different coupling between the cantilever motion and the SQUID for the two measurements. The different coupling is the result of a slightly different position of the cantilever with respect to the flux detector. Using these conversion factors, we find a displacement noise floor at 3 kHz below 10 pm/$\sqrt{\mathrm{Hz}}$ for both spectra.

At frequencies below 5 Hz, the measured noise of the spectrum with vibration isolation becomes larger than that without isolation. However, the amplitude of the vibrations in this frequency range is independent of the coupling between the cantilever motion and the SQUID, indicating that these peaks are not caused by tip-sample movement. Instead, we attribute them to microphonics due to motion of the wiring going to the experiment between the mixing chamber and the top mass of the vibration isolation. The low-frequency motion of the mass-spring system can be removed by using a properly designed damper in parallel to the weak spring, as was shown in Fig. \ref{figure:TF}. This damper would suppress internal resonances of the vibration isolation, for which we expect undamped Q-factors ranging from 100 to 1000, and thereby reduce the microphonics-induced noise.

In the presented experiment, a damper was not implemented for two reasons. First, the power dissipated by the damper would heat the mixing chamber of the cryostat, and thereby reduce the base temperature of the experiment. Secondly, the most commonly used damper at low temperatures is based on the induction of eddy currents by moving a magnet near a conductor. Due to the high sensitivity of our SQUID-based detection for fluctuating magnetic fields, a magnetic damper would deteriorate the detection noise floor in the MRFM experiments. We therefore settled for the internal damping in the weak spring, which is obviously sub-optimal.

\subsection{Cantilever temperature and frequency noise}
\begin{figure}
	\centering
	\includegraphics[width=\columnwidth]{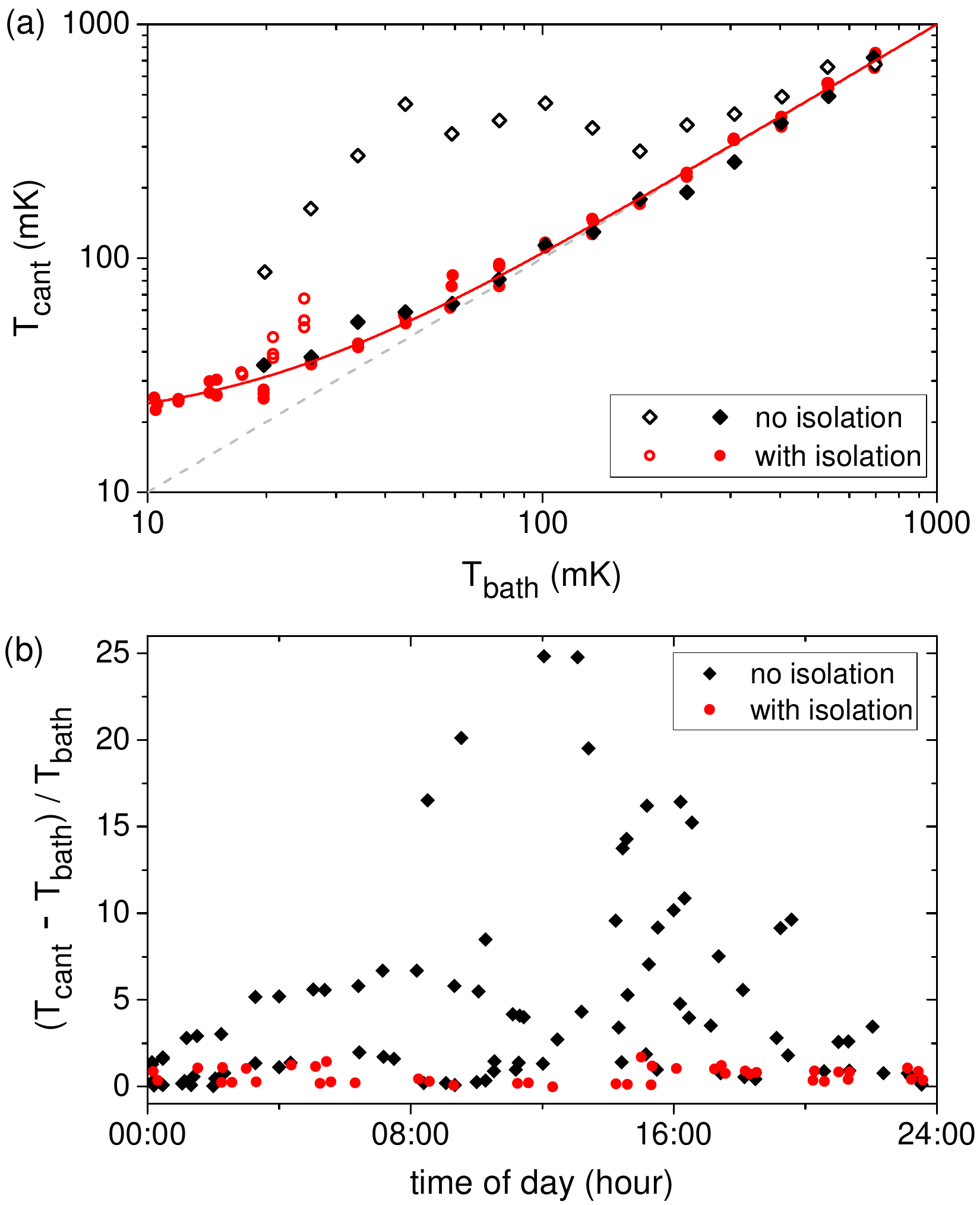}
	\caption{(a) Measurement of the effective cantilever temperature for various bath (bottom mass) temperatures. The black diamonds are data measured without vibration isolation, where the solid diamonds are measured during the night, and the open diamonds during the day. The red circles indicate the measured cantilever temperatures with proper vibration isolation. The open circles are measurements with an elevated cantilever temperature, as explained in the main text. (b) Deviation of the cantilever temperature from the bath temperature plotted against the time of day when the measurement was done. Only bath temperatures below 100 mK are considered. The black diamonds indicate the measurements without vibration isolation. The red circles were measured with vibration isolation.}
	\label{figure:Cant_Temp}
\end{figure}

To further verify the effectiveness of the vibration isolation, we have measured the effective cantilever temperature, following the procedure outlined by \citeauthor{usenko2011} \cite{usenko2011}
Any excitation of the cantilever besides the thermal excitation increases the motional energy of the cantilever to values larger than the thermal energy of the surrounding bath, in our case the bottom mass of the vibration isolation. To measure this effective cantilever temperature, we vary the temperature of the bottom mass between 10.5 mK and 700 mK. At every temperature, we take thermal spectra of the cantilever motion. Using the equipartition theorem, we can derive an effective cantilever temperature from the integrated power spectral density \cite{aspelmeyer2014}:
\begin{equation}
k_B T_{eff} = k_0 \langle x^2 \rangle = k_0 \int_{f_1}^{f_2} \left( S_x - S_0\right) ~ df,
\end{equation}
where $f_1$ and $f_2$ define a small bandwidth around the cantilever resonance frequency, $S_0$ the background determined by the SQUID noise floor, and $S_{x} = c^2 S_{V}$, with $c^2$ the conversion factor between SQUID voltage and cantilever motion. In effect, we calculate the area of the cantilever peak, since this is proportional to the mean resonator energy and thereby the temperature. The resulting cantilever temperature as a function of the bath temperature for the two configurations with and without vibration isolation are shown in Fig. \ref{figure:Cant_Temp}(a). We calibrate the data by assuming that $T_{eff} = T_{bath}$ for the four highest temperatures, where $T_{bath}$ is the temperature of the bottom mass.\\ 

Without the vibration isolation, we observe a large spread in the measured cantilever temperatures. The black data in Fig. \ref{figure:Cant_Temp}(a) shows an example of two data sets, one taken during the night with low effective temperatures and one taken during the day, where the cantilever temperature is increased. As expected, vibrations are most detrimental at low bath temperatures. Figure \ref{figure:Cant_Temp}(b) shows the deviation of the effective temperature from the bath temperature depending on the time of the day (for bath temperatures below 100 mK). The measured effective temperatures show a clear day-night cycle. During the day, the distribution of measured values is much broader than one would expect purely based on the statistical fluctuations of the thermal cantilever energy.  In the worst cases, the effective cantilever temperature can exceed 1.5 K, which corresponds to an equivalent cantilever motion of 0.5 nm. 

When using the vibration isolation, the effective cantilever temperature is nearly equal to the bath temperature for temperatures down to approximately 100 mK, as shown by the red data points in Fig. \ref{figure:Cant_Temp}(a). This means that above 100 mK, the cantilever motion is thermally limited without being significantly disturbed by external vibrations. At lower temperatures, we measure effective temperatures that are slightly increased compared to the bath temperature. However, this increase is independent of the time of day at which the spectra were taken. The elevated effective temperatures are probably due to residual vibrations and an increasingly worse heat conductivity at low temperatures. The red line is a fit to the data to a saturation curve of the form $T_{eff}=(T^n+T_0^n)^{1/n}$, where $T_0$ is the saturation temperature, and n is an exponent determined by the temperature dependence of the limiting thermal conductance. We obtain $T_0$ = 19.7 mK and $n$ = 1.5. This saturation temperature implies an improvement of a factor of 75 when compared to the 1.5 K measured at certain times without the vibration isolation, and corresponds to an effective cantilever motion of 60 pm.

When performing the fit, several data points were not taken into account, indicated by open red circles in Fig. \ref{figure:Cant_Temp}(a). Before taking those spectra, measurements at much higher temperatures had been performed and the system had not reached thermal equilibrium yet, leading to higher effective cantilever temperatures.

Please note that we still observe some unwanted resonances close to the cantilever's resonance frequency, as visible in \ref{figure:spectrum}(a). These resonances prevent us from obtaining a reliable cantilever temperature when, due to a shifting cantilever frequency, these resonances start to overlap with the cantilever's resonance frequency. This indicates that there is room for even further improvements.
 
\begin{figure}
	\centering
	\includegraphics[width=\columnwidth]{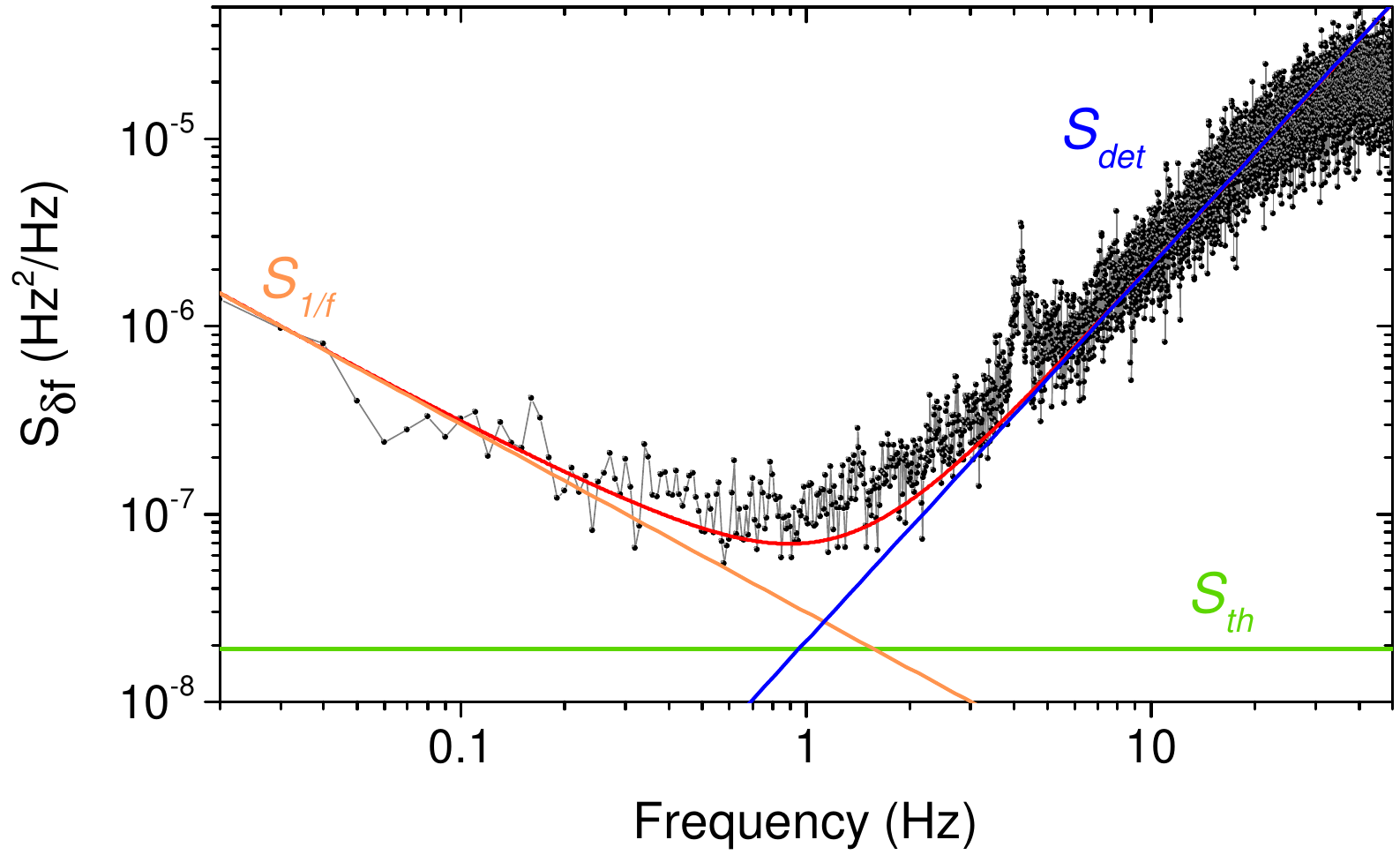}
	\caption{The frequency noise $S_{\delta f}$ of the MRFM cantilever with proper vibration isolation, measured at $T$ = 15 mK. The cantilever is oscillated with an amplitude of 60 nm\textsubscript{rms}. The frequency noise is composed of the detector noise $S_{det}$ (blue), thermal noise  $S_{th}$ (green), and 1/f noise $S_{1/f}$ (orange). The sum of the three is shown in red. The frequency noise floor is found to be 0.3 mHz/$\sqrt{\mathrm{Hz}}$.}
	\label{figure:PLL_Noise}
\end{figure}

For MRFM, the relevance of the low cantilever temperature can be demonstrated by looking at the frequency noise spectrum of the cantilever, as many MRFM protocols are based on detecting minute shifts of the resonance frequency \cite{mamin2007,wagenaar2016}. The frequency noise is measured by driving the cantilever to a calibrated amplitude $A$ = 60 nm\textsubscript{rms}, using a piezoelement. A phase-locked loop (PLL) of a Zurich Instruments lock-in amplifier is used to measure the resonance frequency of the cantilever over time, from which we can calculate the frequency noise spectrum $S_{\delta f}$, which is shown in Fig. \ref{figure:PLL_Noise}. The total frequency noise is given by the sum of three independent contributions \cite{yazdanian2008}: the detector noise $S_{det} = \frac{S_{xn}}{A^2}f^2$ with $S_{xn}$ the position noise, the thermal noise $S_{th} = \frac{k_B T f_0}{2 \pi A^2 k_0 Q}$, and a 1/f noise term $S_{1/f}$. In Fig. \ref{figure:PLL_Noise}, the three terms are indicated by the blue, green, and orange solid lines, respectively, using a cantilever temperature of 15 mK, a measured $Q$ = 20500, and a position noise $\sqrt{S_{xn}}$ = 9 pm/$\sqrt{\mathrm{Hz}}$. The sum of all individual contributions is shown in red. We find a total frequency noise 0.4 mHz in a 1 Hz measurement BW. For a 3000 Hz resonator, this equates to a stability of 0.13 ppm. In typical frequency-shift-based MRFM experiments, the interaction between the cantilever and the spins in the sample induces frequency shifts of several mHz \cite{rugar2004,garner2004,alexson2012,wagenaar2016}. Thus, the current frequency noise floor would allow for single-shot measurements or smaller sample volumes. Due to the relatively low cantilever amplitude and corresponding high detector noise, the detector noise was of similar magnitude to the thermal noise, so a further reduction of the noise floor is possible.

\section{Conclusions}
A mechanical vibration isolation intended for scanning probe microscopy experiments in a cryogen-free dilution refrigerator has been designed and constructed. The vibration isolation offers a large improvement in the measured vibrations in combination with an outstanding thermal conductance between the mixing chamber and the bottom of the isolation, with a base temperature of 10.5 mK at the bottom mass. The high cooling power of 113 $\upmu$W at 100 mK means that a low temperature can be maintained even for experiments where some power dissipation cannot be avoided. The equivalence between electrical and mechanical filters offers a simple and convenient approach to precisely calculate all properties of a mechanical low pass filter in the design phase. The theory shows a large tolerance for the exact mechanical properties of all components, allowing for tailoring of the system to various environments.

Measurements of the effective temperature of a soft mechanical resonator indicate that an effective cantilever temperature of about 20 mK can be achieved. This combination of minimal vibrational noise and low energies in the resonator opens up the possibility for exciting experiments, for instance testing models of wave-function collapse \cite{bassi2013,vinante2016,kleckner2006}, as well as scanning probe investigations of materials showing exotic behaviour at very low temperature.

Furthermore, the ultralow frequency noise achieved using our new vibration isolation can be used for even more sensitive frequency-shift-based MRFM protocols, in which the coupling between the resonator and spins in the sample induces minute changes in the effective stiffness, and thereby the resonance frequency \cite{garner2004,chen2013,wagenaar2016}. The lower cantilever effective temperature directly translates to a lower thermal force noise in the cantilever, given by $S_{th} = 4 k_B T \gamma b$, with $\gamma$ the damping of the resonator and $b$ the measurement bandwidth. For the experimental parameters described in Sec. \ref{sec:VibExp} and the measured cantilever temperature of 20 mK, we find a force noise $\sqrt{S_{th}} \lesssim$ 500 zN/$\sqrt{\mathrm{Hz}}$. This extreme force sensitivity would allow for the MRFM detection volume to be scaled down more and more towards single nuclear spin resolution.

\section{Acknowledgments}
The authors thank M. Camp, F. Schenkel, J. P. Koning, and L. Crama for technical support. The authors thank W. L\"{o}ffler for valuable discussions and proofreading the manuscript. We thank G. L. van der Stolpe and F. G. Hoekstra for proofreading the manuscript. We acknowledge funding from the Netherlands Organisation for Scientific Research (NWO) through a VICI fellowship to T. H. O, and through a VICI fellowship to D. B.

\bibliography{Bibliography_VI}
\bibliographystyle{apsrev4-1}

\end{document}